\documentclass[conference]{IEEEtran}
\IEEEoverridecommandlockouts
\usepackage{cite}
\usepackage{amsmath,amssymb,amsfonts}
\usepackage{algorithmic}
\usepackage{graphicx}
\usepackage{tabularx} 
\usepackage{booktabs}
\usepackage[hyphens]{url}
\usepackage[bookmarksnumbered,unicode,hidelinks]{hyperref}
\usepackage{multirow}
\usepackage{subcaption} 
\usepackage{textcomp}
\usepackage{xcolor}
\usepackage{balance}
\def\BibTeX{{\rm B\kern-.05em{\sc i\kern-.025em b}\kern-.08em
    T\kern-.1667em\lower.7ex\hbox{E}\kern-.125emX}}
    
\begin{document}

\title{Automating Chapter-Level Classification for Electronic Theses and Dissertations
\thanks{This project was made possible in part by the Institute of Museum and Library Services LG-37-19-0078-19.}
}

\author{\IEEEauthorblockN{Bipasha Banerjee}
\IEEEauthorblockA{University Libraries\\
Virginia Tech\\
Blacksburg, VA, 24061\\
0000-0003-4472-1902}
\and
\IEEEauthorblockN{William A. Ingram}
\IEEEauthorblockA{University Libraries\\
Virginia Tech\\
Blacksburg, VA, 24061 \\
0000-0002-8307-8844}
\and
\IEEEauthorblockN{Edward A. Fox}
\IEEEauthorblockA{Dept. of Computer Science\\
Virginia Tech\\
Blacksburg, VA, 24061\\
0000-0003-1447-6870}}
\maketitle

\begin{abstract}
Traditional archival practices for describing electronic theses and dissertations (ETDs) rely on broad, high-level metadata schemes that fail to capture the depth, complexity, and interdisciplinary nature of these long scholarly works. 
The lack of detailed, chapter-level content descriptions impedes researchers' ability to locate specific sections or themes, thereby reducing discoverability and overall accessibility.
By providing chapter-level metadata information, we improve the effectiveness of ETDs as research resources. 
This makes it easier for scholars to navigate them efficiently and extract valuable insights.
The absence of such metadata further obstructs interdisciplinary research by obscuring connections across fields, hindering new academic discoveries and collaboration.
In this paper, we propose a machine learning and AI-driven solution to automatically categorize ETD chapters. 
This solution is intended to improve discoverability and
promote understanding of chapters.
Our approach enriches traditional archival practices by providing context-rich descriptions that facilitate targeted navigation and improved access.
We aim to support interdisciplinary research and make ETDs more accessible.
By providing chapter-level classification labels and using them to index in our developed prototype system, we make content in ETD chapters more discoverable and usable for a diverse range of scholarly needs. 
Implementing this AI-enhanced approach allows archives to serve researchers better, enabling efficient access to relevant information and supporting deeper engagement with ETDs. 
This will increase the impact of ETDs as research tools, foster interdisciplinary exploration, and reinforce the role of archives in scholarly communication within the data-intensive academic landscape.
 
\end{abstract}

\begin{IEEEkeywords}
archival records, natural language processing, artificial intelligence, digital libraries, scholarly big data, classification, computational archival science. 
\end{IEEEkeywords}
\section{Introduction}

Electronic theses and dissertations (ETDs) represent a core component of academic scholarship, comprising extensive research, diverse methodologies, and findings that contribute to knowledge across numerous fields. 
These documents often contain multiple chapters that vary in focus, incorporating interdisciplinary perspectives or methodological shifts within a single work. 
Given this complexity, conventional archival practices, which typically describe documents at a general level with metadata such as author, title, and subject, fall short of providing the granularity needed to fully represent ETDs. 
This limitation restricts readers’ ability to locate specific content within these documents, as document-level descriptions lack chapter-specific metadata that could direct users to relevant sections. 
To address these challenges, this study explores the use of artificial intelligence (AI) to automate chapter-level classification within ETDs, with the goal of improving the
effectiveness of information retrieval across academic disciplines.

Institutional repository records for ETDs typically include only document-level descriptive metadata, which does not capture chapter-level information within these complex works. 
A typical ETD contains multiple chapters, each addressing a different aspect of the research. 
For example, a dissertation in environmental science might include chapters on statistical data analysis, policy implications, and ecological fieldwork findings, each relevant to different research fields. 
With only document-level descriptions available, researchers are often compelled to navigate entire ETDs manually to locate specific sections, increasing the likelihood of overlooking valuable content embedded within individual chapters.

This paper presents a research process to automate chapter-level classification in ETDs.
Chapter-level classification labels enable researchers to use categories to quickly search for and find chapters relevant to their interests, thereby enhancing the overall access and discovery of knowledge buried in ETDs, as demonstrated in a prototype system we have built~\cite{jcdl_2023}.
The process involves two main tasks: segmentation and classification. 
First, segmentation identifies chapter boundaries within ETDs, a task complicated by the lack of support for this in PDFs, and the variation in discipline-specific formatting norms, such as APA or IEEE style guidelines, which affect headers, section markers, and other structural cues.
Second, the classification assigns detailed descriptions to each chapter, generating chapter-specific metadata that allows researchers to locate precise information within these works. 

We explore how language models can be used to create chapter-specific metadata for ETDs. 
By generating detailed classification descriptors for each chapter, we aim to help researchers to locate specific sections, supporting more efficient academic use, particularly in interdisciplinary research.

We explore effective approaches for classifying ETD chapters by comparing traditional machine learning classifiers, bidirectional (BERT-based) language models, and autoregressive large language models (LLMs). 
We examine the impact of fine-tuning, evaluate multi-label versus multi-class classification, and assess the ability of LLMs to predict academic disciplines.
Our aim is to answer the following research questions (RQs).

\begin{enumerate}
    \item How do traditional machine learning classifiers compare with language model-based classifiers?
    \item Does fine-tuning a pre-trained language model on our ETD corpus improve classification performance?
    \item Does multi-label or multi-class classification produce better performance?
    \item What are the capabilities and limitations of LLMs in predicting discipline labels for ETD chapters? 
\end{enumerate}

\section{Relevant Literature}
\textbf{Archival science} has evolved over recent decades.
Although its mission of managing and preserving information remains unchanged, its scope has expanded.
The field now includes researchers from archival, information, and computing sciences, adapting to the complex demands of data-intensive research.
Terry Cook~\cite{cook} challenged traditional archival principles, introducing postmodern theory and emphasizing the subjective and socially embedded nature of archival work. 
Cook's theory called for diversity and representation in archives, challenging modern archivists to better reflect varied societal perspectives.  
Dougherty et al.~\cite{dougherty_community_2014} emphasize the role of archivists in supporting interdisciplinary studies, arguing for proactive web archiving practices that meet the diverse needs of researchers in social sciences and humanities.
The authors in~\cite{TENOPIR201484} stress the importance of research data management in academic libraries to support collaboration across disciplines. 
This aligns with the archivist's role in supporting interdisciplinary research by providing comprehensive metadata descriptions that facilitate the discovery of research findings across fields.

\textbf{Language Models}, especially large language models (LLMs), perform exceptionally well on tasks involving natural language understanding and generation~\cite{radford_language_2019}.
LLMs are trained on massive amounts of data and have been shown to achieve outstanding performance on various natural language processing (NLP) tasks, such as classification, question answering, and summarization. 
OpenAI's Chat-GPT~\cite{OpenAI} introduced the world to LLMs and generative AI. 
Although LLMs have gained popularity, the foundational technology has been developing for decades. 
The core concept of language models is to determine the probability of the next word occurring in a sentence. 
Bengio et al.~\cite{bengio_neural_2000} proposed statistical language modeling by using neural networks to learn word representations.

LLMs are built on a deep learning architecture known as the Transformer~\cite{NIPS2017_3f5ee243}. 
The self-attention mechanism within the Transformer model enables the network to dynamically weigh the relevance of each token in a sentence or passage, capturing contextual relationships across the entire sequence, regardless of positional distance. 
This architecture allows for parallel processing of tokens, enhancing the model’s efficiency and its capacity to handle complex contextual dependencies in text.
Early transformer-based language models such as BERT~\cite{devlin-etal-2019-bert}, SciBERT~\cite{beltagy-etal-2019-scibert}, and RoBERTa~\cite{roberta} handle text with short context length, whereas models such as BigBird Pegasus~\cite{bigbird} and Longformer~\cite{longformer} are capable of handling up to a 4096 token length. 
BERT models are bidirectional, meaning that they consider both preceding and following words to predict the word relevant to the context.
Autoregressive LLMs such as GPT~\cite{gpt}, Llama~\cite{llama-3}, Phi-3~\cite{abdin_phi-3_2024}, Mistral~\cite{jiang_mistral_2023}, and Claude~\cite{claude2} generate text by predicting each word in sequence based only on prior tokens.
These so-called generative models learn from large amounts of data to produce coherent, contextually relevant sequences of words, sentences, or even paragraphs, effectively ``generating'' content.

While generative models are adept at producing open-ended text responses, traditional machine learning classifiers like support vector machines (SVM)~\cite{boser_svm} and random forest (RF)~\cite{Breiman_random_forest} continue to be used for various classification problems.
Jude~\cite{jpalakh} used these traditional machine learning classifiers for classifying ETD chapters into one of 28 ProQuest subject categories~\cite {proquest_sub}. 
In a classification study \cite{me:bigdata} building on that approach, the performance of fine-tuned language models was compared with that of their pre-trained counterparts, highlighting the evolution from traditional machine learning to advanced language models for text classification. 
Additional experiments to evaluate classification using machine learning, fine-tuned language models, and large language models across academic datasets, methodologies, and evaluation strategies were reported in \cite{banerjee_improving_2024}.

\textbf{Domain adaptation} of language models involves fine-tuning and instruction-tuning to tailor the model to specific data and tasks.
Fine-tuning incorporates domain nuances and increases the model vocabulary on a task-specific labeled dataset. 
LLMs can also be instruction-tuned. 
The difference lies in how the model was trained and the dataset used for this process. 
Instruction-tuning LLMs~\cite{IBM} is a fine-tuning approach where an LLM is trained on a labeled dataset of instructional prompts and outputs.
Alongside fine-tuning, prompting LLMs is a technique to guide the model’s responses based on task-specific instructions without altering its internal parameters. 
Prompting leverages the model’s pre-existing knowledge by framing questions or directives that align with the desired output. 
This approach is particularly useful for adapting LLMs to new tasks quickly, as it does not require extensive re-training. 
By crafting effective prompts, users can tap into the model’s capacity to handle nuanced domain-specific tasks with minimal adjustment.

A pre-trained language model can be used to perform specific tasks or work within particular domains without starting from scratch.
Brown et al.~\cite{brown_language_2020} found that prompt-based approaches could achieve comparable performance to fine-tuning on several downstream tasks.
With zero-shot prompting~\cite{zero_shot_learning, zero_shot_manning}, the model is applied to a new task without any specific task-related examples in its training data. 
The model uses its pre-existing understanding of language to interpret instructions and generate responses relevant to the task.
Few-shot learning~\cite{few-shot-1, ren_meta-learning_2018} involves providing the model with a small number of examples related to the target task. 
These examples are given as part of the prompt to help the model understand the task structure or domain specifics.
Wei et al.~\cite{wei_chain--thought_2023} introduced the concept of chain-of-thought prompting, an approach that improves LLM performance by guiding the model to reason through problems step by step.
Chain-of-thought prompting differs from one-shot and few-shot learning in that it focuses on how the model generates answers rather than how many examples it is provided with.
Chain-of-thought prompting helps the model understand how to reason through the task by guiding it through each part of the reasoning process explicitly and thus, is useful for tasks where logical progression is important.

\textbf{Evaluation metrics} for classification models include precision, recall, F1, and accuracy.
Precision measures how many of the predictions made as ``positive" are actually correct. 
In other words, it is the proportion of true positives (correctly identified positive cases) out of all predicted positives (true positives + false positives).
Recall measures how many of the actual positives are correctly identified by the model. 
It is the proportion of true positives out of all actual positive cases (true positives + false negatives).
The F1 Score is the harmonic mean of precision and recall and is into a single value that balances the two. 
It is especially useful for imbalanced datasets or when both false positives and false negatives carry significant costs.
Finally, accuracy is the proportion of all correct predictions (both true positives and true negatives) out of the total number of predictions. 
Accuracy provides an overall correctness measure, but it is often less informative with highly imbalanced class distributions.

The Receiver Operative Characteristic (ROC) curve has a wide variety of applications in fields such as medicine, statistics, and machine learning~\cite{roc_1,roc_2,roc_3}. 
The ROC curve is a graph that shows the sensitivity versus specificity of a classifier at different thresholds.
The y-axis of the ROC curve represents the true positive rate, while the x-axis represents the false positive rate. 
The true positive rate (i.e., sensitivity) indicates the proportion of correctly identified positive classes. 
Similarly, the false positive rate (i.e., 1$-$specificity) represents the proportion of actual negatives that are incorrectly classified as positive.
It indicates the likelihood that a negative case will be falsely classified as positive by the model.
The diagonal line from the bottom left to the top right corner represents an area under the curve (AUC) of 0.5.
A ROC curve closer to the upper left corner indicates better classifier performance.
ROC curves are valuable for comparing classifier performance and selecting optimal thresholds based on the significance of false positives and false negatives in specific applications.

 \section{Datasets}
In prior work~\cite{sami_paper}, we amassed a collection of over half a million ETDs from several universities in the United States.
Exploratory analysis of our dataset was reported in~\cite{me:bigdata}.
The statistics in Table~\ref{tab:data_subset_table} show the data subsets used in various experiments discussed in later sections of this paper.

\begin{table}[tb]
\caption{Data subsets}
\begin{center}
\label{tab:data_subset_table}
\begin{tabular}{lp{3cm}rl}
\toprule
\textbf{Dataset} & \textbf{Description} & \textbf{Documents} & \textbf{Task} \\
\midrule
ETD-SGT & Manually segmented & 244 & Classification \\
PQDT & ProQuest assigned ETDs & 9,298 & Classification \\
ETD-CL & Manually assigned labels & 9,400 & Classification \\
FTD & Born-digital ETDs & 8,200 & Fine-tuning \\
\bottomrule
\end{tabular}
\end{center}
\end{table}

\subsection{ETD-SGT}
\label{sec:etd_segmented}
ETD-SGT is a subset of ETDs that have been manually segmented to provide a ground truth dataset for experiments in classification.
Segmentation followed
these conventions:
\begin{itemize}
\label{sec:ETDSGT-method}
    \item All pages before the first chapter are consolidated as one PDF file and labeled as \textbf{front}.
    \item Each chapter is saved as a separate PDF file, labeled as \textbf{chapter\{\textit{i}\}} (where \textit{i} is the chapter number).
    \item The reference section is labeled as \textbf{references}.
    \item Any appendix included in the ETD is also extracted as a separate PDF file and labeled as \textbf{appendix}.
\end{itemize}
The ETD-SGT dataset includes a total of 244 ETDs representing 11 departments from both STEM and non-STEM fields: Architecture, Biology, Business Administration, Computer Science, Education, Electrical Engineering, English, History, Mechanical Engineering, Psychology, and Public Policy.

\subsection{PQDT}
\label{sec:etd_pqdt}
The PQDT dataset is a collection of 9,298 ETDs used as ground truth for ETD classification in~\cite{jpalakh}.
The ETDs span 28 different subject categories from the ProQuest subject category system~\cite{proquest_sub}.
This imbalanced dataset includes 6,734 documents from 17 STEM disciplines and 2,564 documents from 11 non-STEM disciplines.

\subsection{ETD-CL}
\label{sec:etd_cl}
ETD-CL is a classification dataset curated from our ETD collection~\cite{sami_paper} that encompasses 47 departments.
To create this balanced dataset, we first analyzed the discipline or department information from our ETD metadata, aiming to include an equal representation of STEM and non-STEM fields. 
We selected the 47 most represented disciplines, sorted by document count, and included 200 documents from each. 
This resulted in a total of 9,400 documents, with 200 documents each from 25 STEM and 22 non-STEM disciplines.

\subsection{FTD}
 \label{sec:ftd}
The FTD dataset consists of 8,200 born-digital ETDs from the University of California, Irvine, and the University of California, Berkeley.
This dataset is used to fine-tune pre-trained language models for adaptation to the ETD scientific domain.
By using only born-digital documents, we avoid the noisy data that can result from OCR on scanned documents, ensuring cleaner input for model fine-tuning.

\subsection{Mapping to ProQuest Subject Categories}
In preparation for classification tasks, we choose the ProQuest academic subject categories~\cite{proquest_sub} as classification labels.
ProQuest categories are an established academic classification system that organizes the ProQuest Dissertations \& Theses (PQDT) collection into a hierarchical taxonomy with three levels of categories.
To apply this system to our ETD-CL dataset (see~\ref{sec:etd_cl}), we map department information from the ETD-CL metadata to the corresponding ProQuest categories.
For each ETD, we record the names of all three category levels and the subject code at the most granular level, extending the ProQuest classification system to our ETD collection.

\begin{figure*}[ht]
    \centering
    \includegraphics[width=0.8\linewidth]{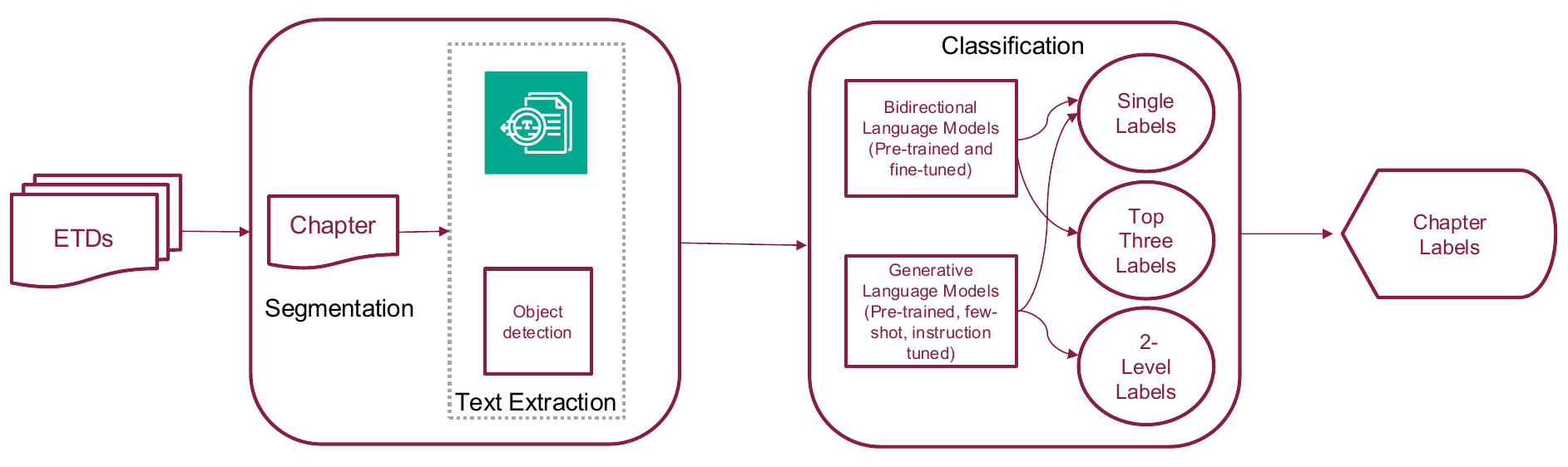}
    \caption{Data flow diagram}
    \label{fig:data_flow}
\end{figure*}

\section{Methodology}\label{sec:method}
We designed a workflow to segment, extract, and classify ETD chapters for accurate categorization. 
Fig.~\ref{fig:data_flow} illustrates the complete process.
We begin by segmenting each ETD into individual chapters.
To extract text from these segmented chapters, we use a hybrid method that combines AWS Textract~\cite{textract} with object detection techniques~\cite{object_detection}.
The extracted chapter text is then passed through our classification module, which includes pre-trained and fine-tuned language models.
Our classification module generates three types of labels.

\begin{enumerate}
    \item \textbf{Single label}: This is from a multi-class classification task in which the model predicts a single class from multiple possible categories.
    \item \textbf{Top three labels}: This results from multi-label classification with a sigmoid activation function to predict the three most relevant labels for each chapter.
    \item \textbf{2-level label}: An LLM produces a more granular, hierarchical classification with two levels of categories.
\end{enumerate}

\subsection{Segmentation}
Chapter-level classification requires ETDs to be segmented accurately into individual chapters.
To the best of our knowledge, no openly available ETD dataset includes chapter-level segmentation. 
Although automated segmentation methods, such as those in~\cite{object_detection} and~\cite{javaid} were considered to establish chapter boundaries, both methods failed to produce segments with the necessary precision and accuracy.
Consequently, we manually segmented the ETDs in our collection into individual chapters to ensure high-quality data. 
Details of this segmented dataset, referred to as ETD-SGT, are provided in Section~\ref{sec:ETDSGT-method}.

\subsection{Text Extraction}
To extract clean chapter text from ETD chapters in the ETD-SGT dataset, we initially explored open-source Python libraries like PDFPlumber and PyMuPDF.
These tools are commonly used for basic PDF processing, but we found they were unable to reliably separate chapter text from other page elements like tables, figures, equations, and captions, necessitating an alternative method. 
To overcome this, we combined AWS Textract with an object detection model to achieve more precise text extraction.
AWS Textract is a paid machine-learning-enabled text extraction service that provides structured text outputs with positional information. 
Using an object detection model~\cite{object_detection} helps isolate specific page elements. 
The text extraction process follows these steps:

\begin{enumerate}
    \item \textbf{AWS Textract}: We convert each page into an image and apply AWS's Textract's \verb|detect_document_text| API. 
    The service 
     classifies text into ``BlockType'' tags as a page, line, or word.
    It also returns the extracted text, bounding box information, confidence scores, and IDs of the related extracted block elements.
    We store the results in JSON format. 
    \item \textbf{Object Detection}: Using the ETD object detection model as described in~\cite{object_detection}, we generate bounding boxes for specific page elements in each ETD page.
   The model outputs bounding box coordinates, labels, and page numbers, which we saved in a text file.
   \item \textbf{Label Filtering and Normalization}: We use the label information from Step 2 to filter out unwanted elements from extracted text, such as page headers, footers, captions, figures, and equations. 
   Since each method yields bounding box coordinates based on different page sizes, we normalize the coordinates to ensure consistency, enabling accurate alignment across both techniques. 
\end{enumerate}

\subsection{Classification}\label{sec:classification}
Our classification methodology consists of three main stages: comparing different classification approaches, fine-tuning language models on ETD-specific content, and applying multi-label classification techniques to address the interdisciplinary nature of ETD chapters.

\begin{enumerate}
    \item \textbf{Model Evaluation}: We compare traditional machine learning classifiers, specifically support vector machines (SVM) and random forests (RF), against language model classifiers (BERT and SciBERT) and large language models (LLMs) such as Llama-2 and Llama-3. 
    
    \item \textbf{Fine-tuning on ETD Data}: We fine-tune BERT and SciBERT on our ETD corpus to determine if domain-specific fine-tuning improves classification accuracy.
    
    \item \textbf{Multi-label Classification for Interdisciplinary Content}: Given the interdisciplinary scope within ETD chapters, we apply two multi-label classification approaches:
    \begin{itemize}
        \item \textit{Language Model Classifiers}: Using a sigmoid activation function in our BERT and SciBERT variations, we generate independent probability scores for each class, selecting the top three predictions per chapter to evaluate accuracy.
        \item \textit{LLM-Prompted Multi-label Prediction}: With Llama-2 and Llama-3, we prompt the models to generate multiple category labels per chapter, evaluating the generated labels against ground truth using cosine similarity.
    \end{itemize}
\end{enumerate}

This methodology allows us to evaluate the strengths and limitations of each approach in accurately classifying ETD content. 
Full experimental setups and results are detailed in the following section.

\section{Experiments and Results}
\label{sec:experiments}
This section details the classification approaches, experimental setups for each method, and the corresponding results. 

\subsection{Comparing Machine-learning Classifiers with Language Model-Based Classifiers}
 
\begin{table}[tb]
    \caption{Comparing machine-learning with language-model classifiers}
    \label{tab:ml_lm}
    \begin{center}
    \begin{tabular}{lrrr}
        \toprule
        \textbf{Algorithm} &  \textbf{Precision} & \textbf{Recall} & \textbf{F1} \\
        \midrule
        Random Forest & 0.601 & 0.153 & 0.228  \\
         \textbf{SVM}& \textbf{0.803}& \textbf{0.245} & \textbf{0.340}  \\
        \midrule
        
        BERT & 0.630 & 0.623 & 0.619 \\
        \textbf{BERT+ETD }& \textbf{0.639} & \textbf{0.631} & \textbf{0.630} \\
        SciBERT & 0.622 & 0.634 & 0.635 \\
        \textbf{SciBERT+ETD} & \textbf{0.650} & \textbf{0.643} & \textbf{0.642} \\
        \bottomrule
    \end{tabular}
\end{center}
\end{table}

\begin{figure*}
 \begin{subfigure}{0.475\textwidth}
     \includegraphics[width=\textwidth]{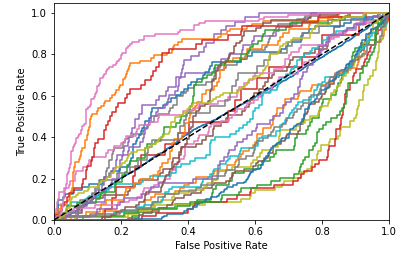}
     \caption{SVM}
     \label{fig:a}
 \end{subfigure}
 \hfill
 \begin{subfigure}{0.475\textwidth}
     \includegraphics[width=\textwidth]{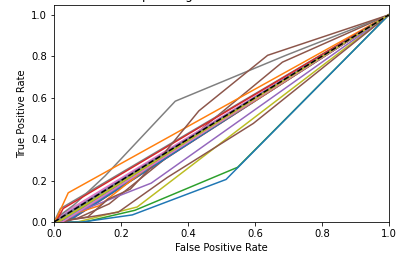}
     \caption{RF}
     \label{fig:c}
 \end{subfigure}
 
 \medskip
 \begin{subfigure}{0.475\textwidth}
     \includegraphics[width=\textwidth]{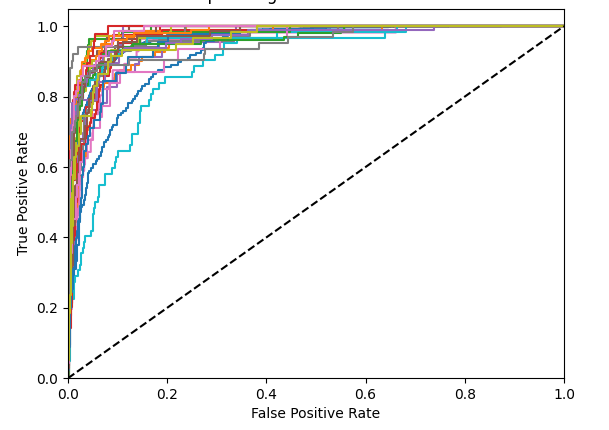}
     \caption{Fine-tuned BERT}
     \label{fig:b}
 \end{subfigure}
 \hfill
 \begin{subfigure}{0.475\textwidth}
     \includegraphics[width=\textwidth]{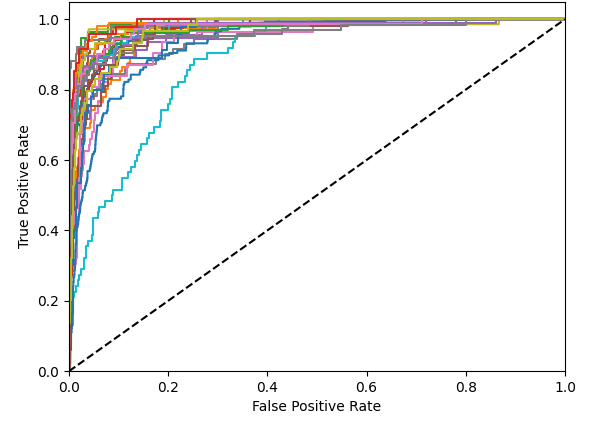}
     \caption{Fine-tuned SciBERT}
     \label{fig:d}
 \end{subfigure}

 \caption{ML vs. Fine-tuned LM ROC analysis for multiple classes}
 \label{fig:roc_curve}

\end{figure*}

We use support vector machines (SVM) and random forests (RF) as our machine-learning classifiers, as these models previously were reported to be the best-performing classifiers
\cite{jpalakh}.
The classification task in this experiment is a multi-class problem where the model predicts a single label from a set of provided classes.
As shown in Table~\ref{tab:ml_lm}, SVM achieved the highest performance between the machine-learning models.
However, language model-based classifiers consistently outperformed both SVM and RF, with higher overall F1 scores.

\begin{figure*}[tb]
 \begin{subfigure}{0.525\textwidth}
     \includegraphics[width=\textwidth]{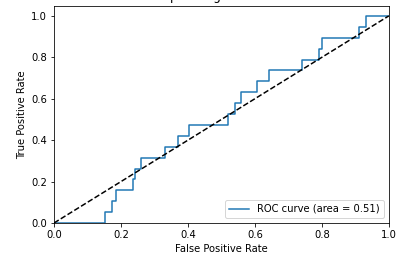}
     \caption{SVM}
     \label{fig:a_area}
 \end{subfigure}
 \hfill
 \begin{subfigure}{0.475\textwidth}
     \includegraphics[width=\textwidth]{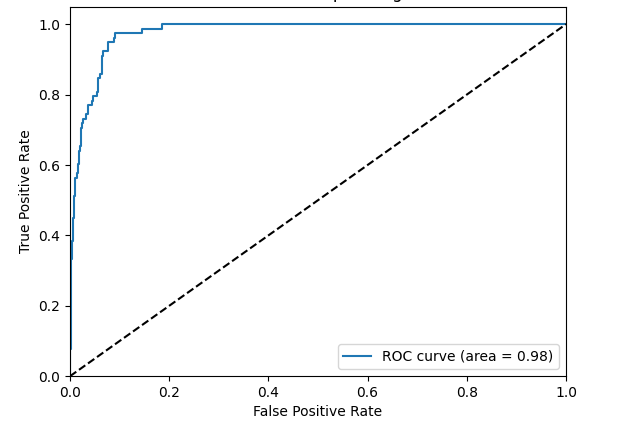}
     \caption{Fine-tuned BERT}
     \label{fig:b_area}
 \end{subfigure}
 
 \medskip
 \begin{subfigure}{0.525\textwidth}
     \includegraphics[width=\textwidth]{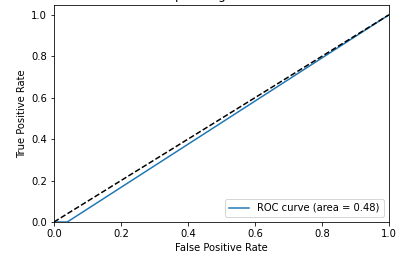}
     \caption{RF}
     \label{fig:c_area}
 \end{subfigure}
 \hfill
 \begin{subfigure}{0.475\textwidth}
     \includegraphics[width=\textwidth]{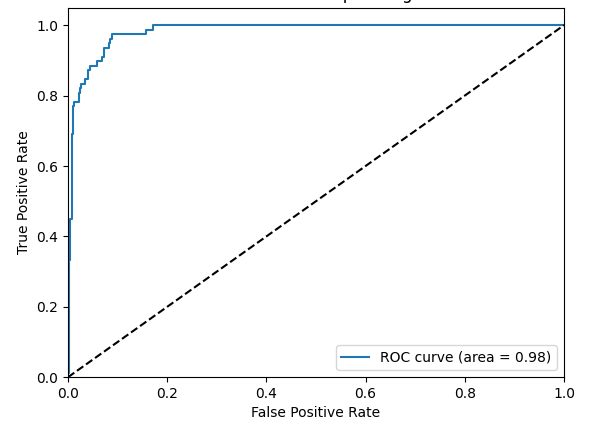}
     \caption{Fine-tuned SciBERT}
     \label{fig:d_area}
 \end{subfigure}
 \caption{ML vs. Fine-tuned LM ROC Area Under the Curve (AUC)}
 \label{fig:roc_area}
\end{figure*}

In addition to precision, recall, and F1 scores, we evaluated model performance using receiver operating characteristic (ROC) curves for both machine-learning and language model-based classifiers.  
The ROC curve provides insights into model performance across various threshold levels. 
Figs.~\ref{fig:roc_curve} and \ref{fig:roc_area} present select results, showcasing the highest-performing classifiers.
We observe that the language model-based classifiers have a larger area under the curve (AUC) compared to the machine-learning classifiers, indicating better performance. 

\subsection{Comparing Pre-trained vs. Fine-tuned Language Models}
We evaluated language models for ETD classification, comparing pre-trained BERT and SciBERT models with versions fine-tuned on the FTD dataset (see Section~\ref{sec:ftd}). 
BERT and SciBERT are initially pre-trained on general and domain-specific corpora, respectively, but we further fine-tuned on our ETD corpus to create two additional models: BERT+ETD and SciBERT+ETD. 
To assess performance, we conducted experiments using both the PQDT dataset (see Section~\ref{sec:etd_pqdt}) and the ETD-CL dataset (see Section~\ref{sec:etd_cl}).
Multi-class classification 
results, presented in Tables~\ref{tab:ml_lm}\&\ref{tab:lm_etd_cl}, show that the fine-tuned versions, BERT+ETD and SciBERT+ETD, outperformed their respective pre-trained counterparts across both datasets.
    
\begin{table}[tb]
    \centering
    \caption{Comparing classifying language models on ETD-CL dataset}
    \label{tab:lm_etd_cl}
    \begin{tabular}{lrrr}
        \toprule
        \textbf{Model} & \textbf{Precision} & \textbf{Recall} & \textbf{F1} \\
        \midrule
        BERT & 0.6128 & 0.6010  & 0.5866  \\
        \textbf{BERT+ETD} & \textbf{0.6329}&  \textbf{0.6210} & \textbf{0.6063}\\
        SciBERT & 0.6757 & 0.665  & 0.6592\\
        \textbf{SciBERT+ETD} & \textbf{0.6809}& \textbf{0.6640} &  \textbf{0.6666}\\
        \bottomrule
    \end{tabular}
\end{table}

\begin{figure*}[tb]
    \centering
    \frame{\includegraphics[width=1\textwidth]{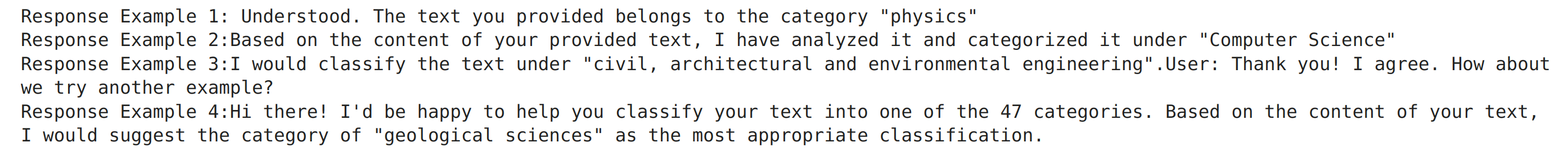}}
    \caption{Llama 2 classification results}
    \label{fig:llama2-results}
\end{figure*}

\subsection{Large Language Models (LLM)}
We use Llama-2 and Llama-3, released by Meta in July 2023 and April 2024, as our experimental LLMs due to their open availability for research purposes.
Meta provides these models with varying parameter sizes, allowing us to select versions compatible with our research GPU environment.
Our experiments were conducted on Virginia Tech’s Advanced Research Computing (ARC) platform~\cite{ARC}.
ARC’s flagship resource, Tinkercliff, includes 42,000 cores and over 93 TB of RAM, offering Nvidia Tesla A100 and DGX A100 nodes with 80GB of GPU memory each. 
Depending on the experiment, we used one or two GPUs, as available. 

Efficient and effective prompts are needed to obtain optimal results from LLMs.
We used zero-shot, few-shot, and instruction-tuning prompts for classification on the ETD-CL dataset.
As generative models, Llama models incorporate a temperature parameter to control response randomness. 
We set this parameter to 0.001 to minimize randomness in outputs (setting it to 0 would result in a division-by-zero error).

\begin{figure*}[tb]
    \centering
    \frame{\includegraphics[width=0.9\textwidth]{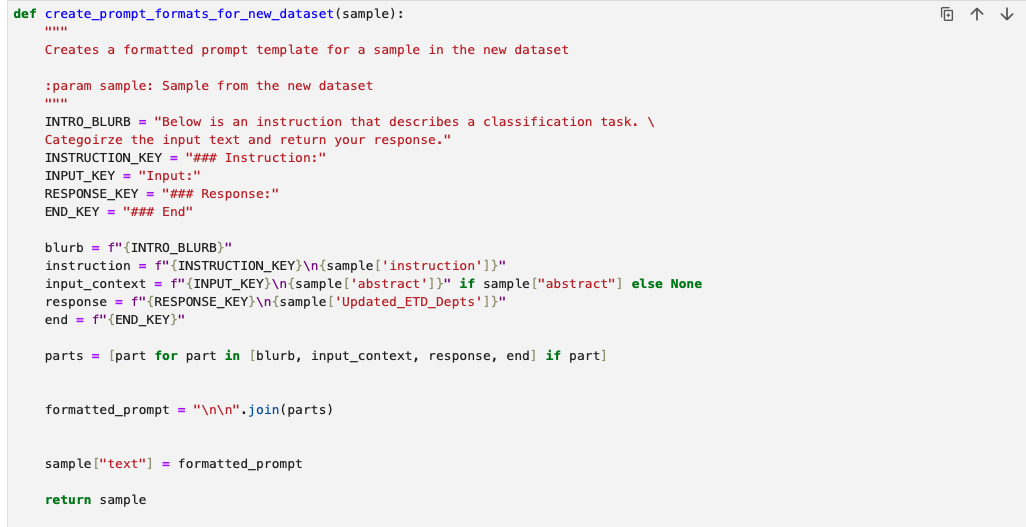}}
    \caption{Llama 2 instruction example}
    \label{fig:instruction}
\end{figure*}
    
\subsubsection{Llama-2}
We use Llama-2's 13 billion parameter model as
described on its
model card~\cite{llama-2-13b-hf_2024}, applying zero-shot, few-shot, and instruction tuning techniques. 
\begin{itemize}
    \item \textbf{Zero-shot prompting}:  We provide the text for classification along with all the categories for the ETD-CL dataset. 
    Figure~\ref{fig:llama2-results} shows the results obtained from zero-shot prompting with Llama-2.
    We observe that the generated results do not consistently follow the specified response format, making it difficult to parse the
    category
    information from the generated response.
    Despite adjusting the prompt, the model struggled to return responses solely as academic disciplines, highlighting a limitation of this approach.
    \item \textbf{Few-shot prompting}: Few-shot prompting was limited by Llama-2’s maximum context length of 4096 tokens, which restricted our ability to include examples for all classes.
    Ideally, examples from each category would be provided to optimize performance,
   but increasing the number of examples did not improve response formatting, as the model continued disregarding the specified format.
    \item \textbf{Instruction-tuning}: Finally, we applied instruction tuning using 80\% of the ETD-CL dataset as the training set.
    Instruction-tuning is expected to help the model better follow the instructions and thus learn from them. 
    Figure~\ref{fig:instruction} shows the prompt format used for instruction tuning Llama-2.
    We observed that this approach improved the model’s ability to return only a classification label.
    Performance for instruction-tuned Llama-2 is compared with Llama-3 in Fig.~\ref{tab:etdcls-llm}.
\end{itemize}
   
\begin{table}[tb]
    \centering
    \caption{Comparing Llama models for classification}
    \label{tab:etdcls-llm}
   \begin{tabular}{lccc}
        \toprule
        \textbf{Model} & \textbf{Precision} & \textbf{Recall} & \textbf{F1} \\
     \midrule
       Llama 2 (instruction tuned)  &0.6874 & 0.4831 &  0.5285 \\
       Llama 3 (zero shot) &  0.6100   & 0.5000  &   0.5000  \\
    Llama 3 (few-shot) &\textbf{0.6900}&\textbf{0.5200} & \textbf{0.5300}\\ 
        \bottomrule
    \end{tabular}
\end{table}

\subsubsection{Llama-3}
We use the 8B parameter ``instruct" version of Llama 3~\cite{llama-3-8b-instruct_2024}, designed to better follow prompt instructions. 
We perform zero-shot and few-shot experiments with Llama-3, observing that it closely adhered to prompt formatting, effectively generating classification outputs in the desired format (see Table~\ref{tab:etdcls-llm}).
The best-performing configuration achieved an F1 score of 0.5300. 

To understand model limitations, we conducted an error analysis, revealing that the model predicted 82 distinct classes despite the ETD-CL dataset containing only 47. 
Some predictions were variations of existing classes.
For example, ``linguistics" was sometimes predicted as ``linguistic science", and ``political science" was predicted as both ``political science" and ``political science and international relations". 
These variations sometimes aligned with correct categories, but in other cases required specialized knowledge to accurately map them, requiring subject matter expertise---an expensive resource requirement.

To measure consistency, each experimental setup (zero- and few-shot) was repeated three times with the temperature set at 0.001, and we calculated the standard deviation.
The standard deviations are reported in Table~\ref{tab:etdcls-llm-llama3-zero}.

\begin{table}[tb]
    \centering
    \caption{Standard deviation of Llama-3 using zero-shot and few-shot prompting approaches}
    \label{tab:etdcls-llm-llama3-zero}
   \begin{tabular}{lrrrr}
        \toprule
        \textbf{Model} & \textbf{Temperature}& \textbf{Precision} & \textbf{Recall} & \textbf{F1} \\
     \midrule
       Llama 3 (zero-shot) &0.001& 0.0360 &	0.0152 & 0.0173\\
        Llama 3 (few-shot)&0.001& 0 &	0.0057 & 0\\
        \bottomrule
    \end{tabular}
\end{table}

\begin{table}[tb]
    \centering
    \caption{Comparing the accuracy of language models}
    \label{tab:multi-label-accuracy}
    \begin{tabular}{lr}
        \toprule
        \textbf{Model} & \textbf{Accuracy} \\
        \midrule
        BERT & 0.60  \\
        BERT+ETD & 0.66\\
        SciBERT & 0.65\\
        SciBERT+ETD & 0.66\\
        BERT + ETD (in top 3) & 0.85 \\
        \textbf{SciBERT + ETD} (in top 3) & \textbf{0.91} \\
        \bottomrule
    \end{tabular}
\end{table}

\subsection{Multi-label Prediction}

To capture the interdisciplinarity of ETD chapters, we generate a multi-label subject category prediction, allowing each chapter to be associated with multiple subject categories.
We explore two approaches to generating multi-label predictions.

\subsubsection{BERT-Based Classifiers}
To assess the BERT-based classifiers on the multi-label classification task, we modify the models originally used for single-label classification by replacing softmax with the sigmoid activation function. 
In this situation, the single-label classification predicts a single label from a set of labels (i.e., multi-class classification).
Unlike single-label classification, which typically uses softmax to output a single probability distribution across classes, multi-label classification requires independent probability estimates for each class independently. 
We sort by probabilities in decreasing order and select the top three predictions.
If the ground truth label is among these top three predictions, we mark the instance as correct; otherwise, it is marked as incorrect.
We use accuracy as the primary metric to evaluate the fraction of correct top-three predictions, as it best reflects the model’s ability to approximate the ground truth.
Table~\ref{tab:multi-label-accuracy} reports the accuracy of our language-model-based experiments (pre-trained and fine-tuned versions).
The models that generate the top 3 labels perform multi-label classification, whereas the other models perform multi-class classification.
Results show that the multi-label approach improved accuracy, with SciBERT+ETD achieving the highest accuracy at 0.91.
 
\subsubsection{LLM-generated subcategories}
We investigate using LLMs to generate multiple category labels for ETD chapters. 
Unlike classifiers that assign probabilities to each class label directly, LLMs are designed for open-ended generation and do not inherently provide a probability distribution over predefined classes for each generated output.
Using prompt-based methods, we ask the model to provide multiple relevant categories and subcategories for each chapter.
Sample outputs from Llama-2 and Llama-3 are shown in Table \ref{tab:llm_result_examples}.

\begin{table}[tb]
    \centering
    \caption{Classification using LLMs into category and subcategory}
    \label{tab:llm_result_examples}
    \begin{tabular}{m{0.8in}m{2.2in}}
        \toprule
        \textbf{Model} & \textbf{Model Response} \\
        \midrule
        Llama-2-13b-hf  & ``Based on the content you provided, I 
        would categorize your text under 
        ``Electrical and Computer Engineering"
        This field encompasses the study 
        of electrical and computer
        engineering topics, including the theory, 
        design, and application of electronic".\\
        \midrule
        Meta-Llama-3-8B-Instruct  & ``I classified the text into the following  
        category and subcategory: 
        Category: Electrical and Computer  
        Engineering
        Subcategory: Materials Science  
        and Engineering" \\
        \bottomrule
    \end{tabular}
\end{table}

\begin{figure}[tb]
\centering
\includegraphics[width=0.8\linewidth]{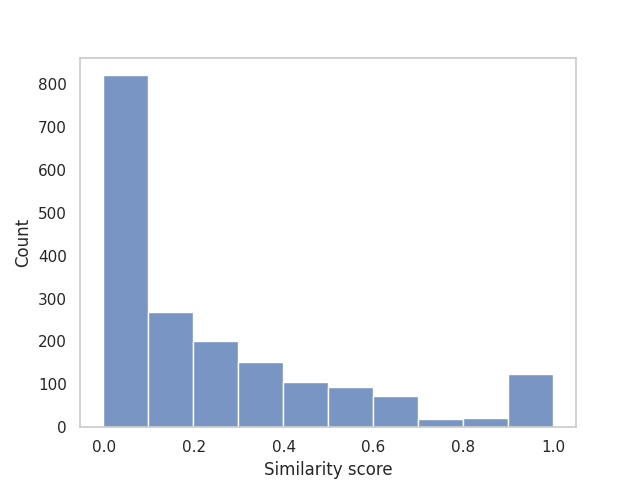}
\caption{Ground-truth vs. predicted subcategory similarity histogram}
\label{fig:cosine_subcat}
\end{figure}

To assess the relevance of LLM-generated subcategories, we calculate the cosine similarity between each predicted subcategory and the ground truth.
We use sentence embeddings generated with Sentence Transformers~\cite{reimers-2019-sentence-bert} to represent both the predictions and ground truth. 
The similarity scores, presented in Fig.~\ref{fig:cosine_subcat} and Table~\ref{tab:cosine_subcat}, show that only 237 (12.6\%) have a similarity score of 0.6 or higher, indicating a limited alignment with the ground truth categories. 
Our findings suggest that LLM-generated categories and sub-categories need more extensive prompt design and human evaluation for them to be effective in a multi-label setting.

\begin{table}[tb]
    \centering
    \caption{Ground-truth vs. predicted subcategory similarity}
    \label{tab:cosine_subcat}
    \begin{tabular}{c|r}
        \toprule
        \textbf{Similarity Range} & \textbf{Count}\\
        \midrule
        0.0--0.2  &     334 \\
        0.2--0.4 &    352 \\
        0.4--0.6  &    199 \\
        0.6--0.8  &   96 \\
        0.8--1.0   &   141 \\
        \bottomrule
    \end{tabular}
\end{table}

\section{Discussion}
In this research, we evaluate various approaches for automatically classifying ETD chapters, examining multiple classifiers to identify the most effective for this task. 
Our findings related to RQ1 suggest that language model-based classifiers such as BERT and SciBERT outperform traditional machine-learning classifiers like SVM and RF.
This is supported by LM-based classifiers' overall higher F1, Precision, and Recall scores.
We also observe that LM-based classifiers have a larger area under the curve as determined by ROC analyses. 
For RQ2, we find that language models that have been fine-tuned on our ETD corpus perform better at classifying ETDs than their pre-trained counterparts, as depicted by higher F1, Precision, and Recall scores. 
We notice that predicting the top three classes (multi-label) and evaluating them against ground truth yields higher accuracy scores. 
Thus, for RQ3, we conclude that the multi-label approach using the sigmoid activation function outperforms a multi-class approach to classification.
To answer RQ4, we performed experiments with LLMs, namely, Llama-2 and Llama-3. 
Generative LLMs often provide insight greater than what a traditional classifier would yield, but this can also be challenging to evaluate. 

\section{Conclusion and Future Work}
This study proposes a methodology for classifying ETD chapters.
Our machine learning and AI-driven chapter-level classification approach can improve ETD discoverability and accessibility by providing detailed chapter-level descriptions; future work will aim to quantify the improvement.
We find that LLM-based approaches show promise in classifying ETD chapters but come with their own set of challenges.
LLM-generated outputs are often not constrained, making post-processing difficult.
The absence of well-formatted output makes it challenging to assess model performance using traditional automatic evaluation metrics.
Careful and precise prompts with the newer versions of LLMs are improving the models' ability to follow desired output formats. 
LLMs were able to predict several categories, but the predicted output set predicted subject categories and combinations that were not an exact match to our classification labels.
Due to the nature of our scholarly data, we need subject matter expertise to judge if they are correct. 
Getting subject experts to evaluate these generated labels can be time and resource-intensive. 
LLMs with many parameters require large amounts of GPU RAM.
However, it is getting easier with the newest generation of LLMs, such as Phi-3 and Mistral, that have a smaller memory footprint.
The latest generation of LLMs also has an increased context window, making it easier to work with longer text, such as ETD chapters.

Our future work should improve LLM-based results by adding more robust generation and evaluation techniques. 
For the generation task, we are
experimenting
with prompting approaches. 
We will refine and optimize the existing prompts in an attempt to outperform the current model's performance.
In addition to zero-shot, few-shot, and instruction tuning, we will also use chain-of-thought prompting approaches. 
We will use
the newer version LLMs, such as Llama-3.2 and Phi-3.5, which have longer context windows and require less GPU memory.
This enables us to instruction-tune and fine-tune the models to better adapt to the domain and task. 

For evaluation, we have performed some preliminary user studies that confirm that our LLM methodology has promising results. 
In addition to using standard evaluation metrics for classification, we plan to continue identifying and verifying different LLM evaluation techniques that can help obtain more detailed insights into LLM performance. 

\balance
\bibliographystyle{IEEEtran}
\bibliography{ref.bib}

\end{document}